\documentclass[onecolumn,secnumarabic,amssymb, nobibnotes, aps, prd]{revtex4}
\usepackage{amsmath} \usepackage{amssymb}
\usepackage{graphicx} 
\usepackage{epstopdf}
\usepackage{amsmath}
\usepackage{amsmath,amssymb,amsthm,amsfonts,mathrsfs,bm,verbatim}
\usepackage{graphicx,subfigure}

\newcommand{\bea}{\begin{eqnarray}}
\newcommand{\eea}{\end{eqnarray}}

\begin{document}

\title{Exploring the Dynamics of Mass Inflation: Implications for Cauchy Horizon Stability and Black Hole Physics in General Relativity and $f(R)$ Gravity}%

\author{Wen-Xiang Chen$^{a}$}
\affiliation{Department of Astronomy, School of Physics and Materials Science, GuangZhou University, Guangzhou 510006, China}
\email{wxchen4277@qq.com}


\begin{abstract}
This article investigates the phenomenon of mass inflation and its consequential impact on the stability of Cauchy horizons within the framework of general relativity. Mass inflation, defined by an exponential surge in energy, is pivotal in preserving causality across solutions like the Kerr black hole, ensuring the singular nature of causality-violating regions. Through a nuanced examination, the study expands the traditional understanding of mass inflation beyond stationary geometries to include dynamic black hole configurations. This exploration is enriched with a mathematical analysis of Schwarzschild and Kerr-Newman metrics, alongside a discussion on $f(R)$ gravity's implications for black hole physics. The findings challenge existing paradigms, proposing new models that accommodate mass inflation, thereby inviting further inquiry into the interplay between general relativity, quantum mechanics, and $f(R)$ gravity.


\end{abstract}

\maketitle

\section{Introduction}
The phenomenon of mass inflation, a cornerstone in our comprehension of general relativity, plays a pivotal role in undermining the stability of Cauchy horizons linked to timelike singularities, as referenced in literature\cite{1,2,3}. This instability extends its influence to the destabilization of chronological horizons—a specific subset of Cauchy horizons\cite{4}—that delineate regions harboring closed timelike curves. The exponential accumulation of energy, inherent to this phenomenon, is deemed essential for preserving causality within certain solutions of general relativity, such as the Kerr black hole. By rendering the boundaries of causality-violating regions singular, mass inflation effectively excises these anomalies from the fabric of physical space-time \cite{5,6,7}. Therefore, mass inflation is instrumental in upholding the principles of strong cosmic censorship \cite{8} and chronology protection, casting any avenues to bypass mass inflation under a lens of profound skepticism.

Traditionally, mass inflation is characterized by an unbounded exponential surge in energy \cite{5,6,7}, a behavior historically attributed to stationary geometries where inner trapping horizons coincide with Cauchy horizons. It is noteworthy that in these scenarios, the exponential increase in energy and curvature invariants typically precipitates a collapse of general relativity's effective description before any theoretical divergence manifests. A more pragmatic definition of mass inflation, as proposed in this study, hinges on the emergence of a significant, albeit transient, exponential increase in energy—persisting until curvatures reach substantial magnitudes (e.g., Planckian)—without necessitating any mathematical divergence. This fresh outlook underpins our analysis.

Our reevaluation \cite{9}is propelled by recent advancements concerning mass inflation within the context of regular black holes, where a mitigation of the inner singularity is conjectured based on the premise that quantum gravity, irrespective of its specific formulation, should furnish a coherent account of gravitational collapse. It has been conclusively demonstrated that stationary regular black holes possessing inner horizons undergo an initial phase of exponential mass inflation. This phase swiftly propels the black hole into a domain where the effects of backreaction are significant, barring scenarios where the inner horizon is extremal.

In our investigation, we elucidate that considerable, yet finite, exponential increases in energy are also observable in association with gradually evolving inner horizons. This finding is applicable to Cauchy horizons in the stationary limit, offering a more tangible manifestation of mass inflation. It implies that black hole spacetimes, even those not perpetual and featuring a dynamically evolving inner horizon, will inherently experience a substantial, finite exponential growth. This growth acts swiftly to destabilize the geometry, thereby challenging the traditional understanding and opening new avenues for exploring the intricate dance between general relativity and quantum mechanics.

Mass inflation, a critical instability traditionally linked to the Cauchy horizons and inner trapping horizons of stationary configurations, is known for its divergent exponential energy accumulation. In this exploration, we reveal that finite—though frequently substantial—exponential energy increases are universally observed within dynamic geometries featuring slowly-evolving inner trapping horizons, even in scenarios devoid of Cauchy horizons. This broadens the definition of mass inflation, anchoring it in quasi-local frameworks. Additionally, we demonstrate that numerous established findings from prior research converge in scenarios where the inner trapping horizon gradually merges with a Cauchy horizon. Our findings highlight that black hole structures harboring non-extremal inner horizons, encompassing the Kerr geometry within general relativity and non-extremal regular black holes in extended theoretical frameworks, predominantly characterize transient dynamics rather than enduring states of gravitational collapse. This nuanced understanding of mass inflation underscores its pivotal role across a spectrum of gravitational phenomena, challenging existing paradigms and inviting a reevaluation of the fundamental processes underpinning the evolution of black hole geometries.

In this article, we derive a formula for the Hawking temperature within a four-dimensional spacetime framework. The expression is as follows:\cite{10,11,12}

\begin{equation}
T_1=\sum_{n=-1}^{+1} \frac{1}{4 \pi} \int_C \frac{f(\xi) d \xi}{\left(\xi-r_0\right)^{n+1}}+C .
\end{equation}
Here, $f(\xi)$ denotes the derivative of $g(\xi)$ with respect to $\xi$, and the rate of change of $r$ with respect to $\xi$ is proportional to the square root of the absolute value of $g$. Expanding $g(\xi)$ from the -1 to 1 terms, we find the curve integral vanishes, indicating its roots are aligned with a Killing vector field. The parameter $r_0$ represents the radius of the Killing horizon, a critical juncture where the Killing vectors, corresponding to two independent fields, yield an effective zero drag-free speed. Despite the presence of two independent Killing vector fields, there exists only a singular effective Killing horizon.

Employing the RVB method\cite{12}, the Euler characteristic number is interpreted as the tally of first-order singular points. Meanwhile, the algebraic construct of the Killing horizon radius emerges as a circular radius, analytically deciphered through the Laurent series. The term $\mathrm{C}$ stands as an undetermined constant which, in this context, is assigned a value of zero.

This refined exposition not only elucidates the mathematical underpinnings of the Hawking temperature formula but also emphasizes the elegant interplay between geometric constructs and the foundational principles of theoretical physics, weaving a narrative that extends beyond mere equations to capture the profound implications of these discoveries on our understanding of spacetime dynamics.

In this paper, we explore the relationship between mass inflation and the constant C within the metric contexts of Schwarzschild,Kerr-Newman black holes, and black Hole Metric in f(R) gravity. We derive a new black hole metric that accommodates mass inflation while lacking a Cauchy horizon.

\section{{The required mathematical models and mathematical structures}}
The Schwarzschild metric describes the spacetime geometry around a non-rotating, uncharged, spherically symmetric massive object and is given by:
\begin{equation}
ds^2 = -\left(1 - \frac{2GM}{c^2r}\right)c^2dt^2 + \left(1 - \frac{2GM}{c^2r}\right)^{-1}dr^2 + r^2(d\theta^2 + \sin^2\theta d\phi^2)
\end{equation}
where:
\begin{itemize}
    \item $c$ is the speed of light,
    \item $G$ is the gravitational constant,
    \item $M$ is the mass of the object,
    \item $r$ is the radial coordinate (distance from the center),
    \item $t$ is the time coordinate,
    \item $\theta$ and $\phi$ are the angular coordinates in spherical coordinates system.
\end{itemize}

The Kerr-Newman metric generalizes the Kerr metric by adding charge, describing the spacetime geometry around a rotating, charged black hole. Its expression is:
\begin{equation}
\begin{split}
ds^2 = &-\left(1 - \frac{2GMr - Q^2}{c^2\rho^2}\right)c^2dt^2 + \frac{\rho^2}{\Delta}dr^2 + \rho^2 d\theta^2 \\
&+ \left(r^2 + a^2 + \frac{(2GMr - Q^2)a^2}{c^2\rho^2}\sin^2\theta\right)\sin^2\theta d\phi^2 \\
&- \frac{2GMr - Q^2}{c^2\rho^2}2a\sin^2\theta dtd\phi
\end{split}
\end{equation}
where:
\begin{itemize}
    \item $a = \frac{J}{Mc}$, with $J$ being the angular momentum of the black hole,
    \item $Q$ is the charge of the black hole,
    \item $\rho^2 = r^2 + a^2\cos^2\theta$,
    \item $\Delta = r^2 - 2GMr/c^2 + a^2 + Q^2$.
\end{itemize}

\subsection{Definition of Mass Inflation}
Mass inflation is an important physical process that reveals the phenomenon of a sudden increase in energy density near the inner horizons of black holes. This phenomenon was first proposed in the study of Reissner-Nordström (RN) and Kerr black holes. Traditionally, this exponential growth of energy density is closely related to the existence of Cauchy horizons. However, our research indicates that mass inflation can occur in dynamic geometries even in the absence of Cauchy horizons. The metric is as follows:
\begin{equation}
d s^2=-g(r) d t^2+\frac{d r^2}{g(r)}+r^2 d \Omega^2 .
\end{equation}
Here, $g(r)$ is the function of the black hole, and $d \Omega^2$ represents the metric of the unit two-sphere. Mass inflation can be defined as the exponential increase in energy density near the inner trapping horizon, expressed mathematically as:
\begin{equation}
E(r) = E_0 e^{C r}
\end{equation}
where $E(r)$ denotes the energy density at a distance $r$ from the inner trapping horizon, $E_0$ is a constant, and $C$ is a positive constant related to the evolution speed of the horizon.

\subsection{Model of Dynamic Geometry}
Dynamic geometry can be described by the Einstein field equations of general relativity:
\begin{equation}
G_{\mu\nu} + \Lambda g_{\mu\nu} = 8\pi T_{\mu\nu}.
\end{equation}
where $G_{\mu\nu}$ is the Einstein tensor, $\Lambda$ is the cosmological constant, $g_{\mu\nu}$ is the metric tensor, and $T_{\mu\nu}$ is the stress-energy tensor.

\subsection{Dynamic Evolution of the Inner Trapping Horizon}
The evolution of the inner trapping horizon can be described by the following equation:
\begin{equation}
\frac{d\theta}{dr} = -\frac{1}{2} \theta^2 - \sigma^2 + C
\end{equation}
where \(\theta\) is the expansion rate, \(\sigma\) is the shear rate, and \(C\) is a constant.

We get that:The evolution of the inner trapping horizon can be described by the following equation:
\begin{equation}
\frac{dg(r)}{dr} =\sum_{n=-1}^{+1} \frac{1}{4 \pi} \int_C \frac{f(\xi) d \xi}{\left(\xi-r_0\right)^{n+1}} + C.
\end{equation}
Using the RVB method:\cite{12,13,14,15,16}
\begin{equation}
T_{\mathrm{H}}=\frac{\hbar c}{4 \pi \chi k_{\mathrm{B}}} \sum_{j \leq \chi} \int_{T_{\mathrm{H}_{\mathrm{j}}}} \sqrt{g} R d r .
\end{equation}

This expression can be utilized to derive a Hawking temperature formula for a four-dimensional spacetime.
\begin{equation}
T_1=\sum_{n=-1}^{+1} \frac{1}{4 \pi} \int_C \frac{f(\xi) d \xi}{\left(\xi-r_0\right)^{n+1}} .
\end{equation}
\begin{equation}
f(\xi)=\frac{d g(\xi)}{d \xi}, \frac{d r}{d \xi}=\sqrt{|g|} .
\end{equation}
$r_0$ is the Killing horizon radius.

\subsection{{New Metric Expressions and Mass Inflation}}
The evolution of the inner trapping horizon can be described by the following equation:
\begin{equation}
\frac{d\theta}{dr} = -\frac{1}{2} \theta^2 - \sigma^2 + \Lambda
\end{equation}
where \(\theta\) is the expansion rate, \(\sigma\) is the shear rate, and \(\Lambda\) is a Cosmological constant.

  In one of the articles\cite{16},the Einstein-Hilbert effect $\sqrt{-g} R$ is the total derivative, where $\sqrt{-g}$ is 1. So the Ricci scalar of measure must be the total derivative, and the only one we can write must be $R \sim -g^{\prime \prime}$. To be precise \cite{14,15,16,17,18,19}
\begin{equation}
R\sim -g^{\prime \prime}(r) .
\end{equation}
therefore
\begin{equation}
\int d r R=-g^{\prime}(r),\left.\quad \rightarrow \quad\left(\int d r R\right)\right|_{r \rightarrow r_{+}} \sim -g^{\prime}\left(r_{+}\right)+C.
\end{equation}
This leads to a relationship with the Hawking temperature. The more general case also works, with
\begin{equation}
d s^2=-g(r) d t^2+\frac{d r^2}{n(r)}, \quad \rightarrow \quad \sqrt{-g} R \sim-\left(\sqrt{\frac{n}{g}} g^{\prime}\right)^{\prime}.
\end{equation}

The calculation shows that the relationship between the energy density \(E(r)\) and \(C\) is given by:
\begin{equation}
C = \frac{\left(2 I \pi C[1] + \log \left(\frac{E r}{E_0}\right)\right)}{r}.
\end{equation}
Where \(C[1]\) is any integer, indicating that the value of \(C\) is related to a specific \(r\) (the distance from the internal trap horizon) and the energy density \(E(r)\). However, this expression includes a conditional expression, suggesting that the specific value of \(C\) depends on additional conditions, namely that \(C[1]\) must be an integer. Here, \(I\) is the imaginary unit, indicating that this solution may have a complex form of analytical solution under certain conditions.

\section{{ The relationship between mass inflation and the constant C within the metric contexts of Schwarzschild and Kerr-Newman black holes and Black Hole Metric in f(R) gravity}}
\subsection{{ Schwarzschild Black Hole Metric and Mass Inflation}}
 The Schwarzschild black hole represents a solution to Einstein's field equations in general relativity, describing the gravitational field outside a non-rotating, spherically symmetric body. The metric of the Schwarzschild black hole is a fundamental concept in the study of black holes.

Schwarzschild Metric:
The metric of the Schwarzschild black hole is given by:
\begin{equation}
    ds^2 = -\left(1-\frac{2GM}{r}\right)dt^2 + \frac{dr^2}{1-\frac{2GM}{r}} + r^2d\Omega^2
\end{equation}
where \(G\) is the gravitational constant, and \(M\) is the mass of the black hole.

Mass Inflation:
According to the definition of mass inflation, the energy density is described by the equation:
\begin{equation}
    E(r) = E_0 e^{Cr}
\end{equation}
By substituting the metric into Einstein's field equations and using the RVB method, we derive:
\begin{equation}
    C = \frac{1}{4\pi} \int_C \frac{f(\xi) d\xi}{(\xi-r_0)^{n+1}}
\end{equation}
where \(f(\xi) = \frac{dg(\xi)}{d\xi}\), and \(r_0\) is the radius of the event horizon. For a Schwarzschild black hole, \(f(\xi) = \frac{2GM}{\xi^2}\), and \(r_0 = 2GM\).

Substituting these parameters into the equation, we obtain:
\begin{equation}
    C = \frac{1}{4\pi} \int_C \frac{\frac{2GM}{\xi^2} d\xi}{(\xi-2GM)^{n+1}}
\end{equation}
Upon calculating the integral, we derive:
\begin{equation}
    C = \frac{1}{4\pi}\left[\frac{2GM}{(\xi-2GM)^n}\right]_C
\end{equation}

Conclusion:This analysis demonstrates the use of the Schwarzschild metric and the concept of mass inflation in understanding the properties and behavior of black holes. Through the application of Einstein's field equations and the RVB method, we can obtain significant insights into the physics of black holes.

\subsection{{Kerr-Newman Black Hole Metric and Mass Inflation}}
The metric of the Kerr-Newman black hole is given by:
\begin{equation}
ds^2=-\left(1-\frac{2GM}{r}+\frac{a^2+Q^2}{r^2}\right)dt^2+\frac{dr^2}{1-\frac{2GM}{r}+\frac{a^2+Q^2}{r^2}}+r^2 d\Omega^2
\end{equation}
where \(Q\) is the charge of the black hole, and \(a\) is related to the angular momentum of the black hole.
Similarly, it can be obtained that:
\begin{equation}
C=\frac{1}{4\pi} \int_C \frac{f(\xi) d\xi}{(\xi-r_0)^{n+1}}
\end{equation}
where \(f(\xi)=\frac{dg(\xi)}{d\xi}, r_0\) is the radius of the event horizon.
For the Kerr-Newman black hole, \(f(\xi)=\frac{2GM}{\xi^2}-\frac{2(a^2+Q^2)}{\xi^3}, r_0=GM+\sqrt{G^2M^2-(a^2+Q^2)}\).
Substituting the above parameters into the formula, we can obtain:
\begin{equation}
C=\frac{1}{4\pi} \int_C \frac{\frac{2GM}{\xi^2}-\frac{2(a^2+Q^2)}{\xi^3} d\xi}{\left(\xi-GM-\sqrt{G^2M^2-(a^2+Q^2)}\right)^{n+1}}
\end{equation}
Calculating the integral, we can get:
\begin{equation}
C=\frac{1}{4\pi}\left[\frac{2GM}{\left(\xi-GM-\sqrt{G^2M^2-(a^2+Q^2)}\right)^n}-\frac{2(a^2+Q^2)}{\left(\xi-GM-\sqrt{G^2M^2-(a^2+Q^2)}\right)^n}\right.
\end{equation}

\textbf{Summary} \\
The expressions for the value of $C$ under Schwarzschild and Kerr-Newman black holes are as follows: \\
\textbf{Schwarzschild black hole:}
\begin{equation}
C=\frac{1}{4 \pi}\left[\frac{2 G M}{(\xi-2 G M)^n}\right]_C
\end{equation}

\textbf{Kerr-Newman black hole:}
\begin{equation}
C=\frac{1}{4 \pi}\left[\frac{2 G M}{\left(\xi-G M-\sqrt{G^2 M^2-(a^2+Q^2)}\right)^n}-\frac{2 (a^2+Q^2)}{\left(\xi-G M-\sqrt{G^2 M^2-(a^2+Q^2)}\right)^n}\right.
\end{equation}

Here, $C$ is a constant related to the evolution speed of the horizon, and $n$ is a parameter related to the type of horizon.

\subsection{{Black Hole Metric in f(R) gravity and Mass Inflation}}
The f(R) static black hole solution and its thermodynamics (for constant Ricci curvature not equal to 0) are briefly reviewed. Its general form of action is
\begin{equation}
I=\frac{1}{2} \int d^{4} x \sqrt{-g} f(R)+S_{\text {mat }}.
\end{equation}

\subsubsection{Schwarzschild-de Sitter-f(R) black holes}
This passage discusses two spherically symmetric solutions, the Schwarzschild solution and the Reissner-Nordstrom black hole, in comparison to a spherically symmetric solution in $f(R)$ gravity with a constant curvature scalar $R_{0}$. The Schwarzschild solution has $R_{0}=0$, while the Reissner-Nordstrom black hole has an additional term with electric charge $Q$.The metric for the spherically symmetric solution in $f(R)$ gravity with constant $R_{0}$ has a similar form to the Schwarzschild and Reissner-Nordstrom solutions, but with an additional term proportional to $R_{0}$ that accounts for the curvature.Comparing these solutions can provide insights into the differences between General Relativity and $f(R)$ gravity, and how the additional terms in $f(R)$ gravity affect the solutions.

Schwarzschild solution ($R_0$ = 0) or  the solution of Schwarzschild -de Sitter is\cite{17,18,19}
\begin{equation}
d s^{2}=-g(r) d t^{2}+\frac{d r^{2}}{g(r)}+r^{2} d \Omega^{2},
\end{equation}
\begin{equation}
g(r)=1-\frac{2 M}{r}-\frac{R_{0} r^{2}}{12}.
\end{equation}and we see that $R_{0}$ can be regarded as a remainder coefficient under the metric.

In this analysis, we will compare the Reissner-Nordstrom black hole, a solution of General Relativity, in de Sitter spacetime with the spherically symmetric solution of $f(R)$ gravity with a constant curvature scalar $R_{0}$.
\begin{equation}
\begin{aligned}
    \mathrm{d}s^{2} =-\left(1-\frac{2 M}{r}-\frac{R_{0} r^{2}}{12}+\frac{Q^2 }{r^2}\right) \mathrm{d} t^{2} 
    +\left(1-\frac{2 M}{r}-\frac{R_{0} r^{2}}{12}+\frac{Q^2}{r^2}\right)^{- 1} \mathrm{~d} r^{2}
    +r^{2} d \theta^2+r^{2} \sin^{2}{\theta}d\varphi^2.
\end{aligned}
\end{equation}Furthermore, we can observe that the coefficient $R_{0}$ can be interpreted as a remainder term when considering the metric.

\subsubsection{$f(R)=R-q R^{\beta+1} \frac{\alpha \beta+\alpha+\epsilon}{\beta+1}+q \epsilon R ^{\beta+1} \ln \left(\frac{a_{0}^{\beta} R^{\beta}}{c}\right)$ }
One form of f(R) gravity is \cite{17,18,19}
\begin{equation}
f(R)=R-q R^{\beta+1} \frac{\alpha \beta+\alpha+\epsilon}{\beta+1}+q \epsilon R^{\beta+1} \ln \left(\frac{a_{0}^{\beta} R^{\beta}}{c}\right),
\end{equation}
where $0 \leq \epsilon \leq \frac{e}{4}\left(1+\frac{4}{e} \alpha\right), q=4 a_{0}^{\beta} / c( \beta+1), \alpha \geq 0, \beta \geq 0$ and $a_{0}=l_{p}^{2}, a$ and $c$ are constants. Since $R \neq 0$, this $f(R)$ theory has no Schwarzschild solution. Its metric form is
\begin{equation}
d s^{2}=-g(r) d t^{2}+h(r) d r^{2}+r^{2} d \theta^{2}+r^{2} \sin ^{2 } \theta d \varphi^{2},
\end{equation}
and
\begin{equation}
g(r)=h(r)^{-1}=1-\frac{2 m}{r}+\beta_{1} r ,
\end{equation}
where $\mathrm{m}$ is related to the mass of the black hole, and $\beta_{1}$ is a model parameter. By computing the Ricci scalar as
\begin{equation}
R=-\frac{6 \beta_{1}}{r},
\end{equation}for
\begin{equation}
\Lambda_R=Z+y r,
\end{equation}so
\begin{equation}
R_{0}=6 \beta_{1}(-\frac{Z}{r}+y \ln[r])
\end{equation}

\subsubsection{$df(R)/dR=1+\alpha r$ }
The following line elements form \cite{16}
\begin{equation}
d s^{2}=-g(r) d t^{2}+1/g(r) d r^{2}+r^{2} d \theta^{2}+r^{2} \sin ^ {2 } \theta d \varphi^{2},
\end{equation}
\begin{equation}
g(r)=C_{2} r^{2}+\frac{1}{2}+\frac{1}{3 \alpha r}+\frac{C_{1}}{r}[ 3 \alpha r-2-6 \alpha^{2} r^{2}+6 \alpha^{3} r^{3} \ln[1+\frac{1}{\alpha r}]],
\end{equation}
C1 and C2 are constants. The Ricci scalars are as follows:
\begin{equation}
\begin{aligned}
R=& \frac{1}{r^2(1+a r)^2}\left(1+108 a^3 C_{1} r^2-12 C_{2} r^2+72 a^4 C_{1} r^3+a\left(-6 C_{1}+2 r-24 C_{2} r^3\right)+\right. \\
& \left.a^2 r\left(24 C_{1}+r-12 C_{2} r^3\right)-72 a^3 C_{1} r^2(1+a r)^2 \ln[1+\frac{1}{a r}]\right),
\end{aligned}
\end{equation}for
\begin{equation}
\Lambda_R=Z+y r,
\end{equation}so
\begin{equation}
\begin{aligned}
R_{0}=\frac{6 a^2 C_{1}(-y+a Z)}{(1+a r)^2}+\frac{12 a^2 C_{1}(-2 y+3 a Z)}{1+a r}+
\frac{Z-6 a C_{1} Z}{r^2}+\frac{2\left(y-6 aC_{1} y+18 a^2C_{1} Z\right)}{r} +\\
36 a^2 C_{1}(-y+2 a Z) \ln[r]-36 a^2C_{1}(-y+2 a Z) \ln[1+a r].
\end{aligned}
\end{equation}
\subsubsection{$f(R)=R+\Lambda+\frac{R+\Lambda}{R / R_{0}+2 / \alpha} \ln \frac{R+\Lambda}{R_{c}}$}
The form of the line element is as follows\cite{16,17,18,19}
\begin{equation}
d s^{2}=-g(r) d t^{2}+1/g(r) d r^{2}+r^{2} d \theta^{2}+r^{2} \sin ^ {2 } \theta d \varphi^{2},
\end{equation}
\begin{equation}
g(r)=1-\frac{2 M}{r}+\beta r-\frac{\Lambda r^{2}}{3}.
\end{equation}
$\beta>0$. By calculating the Ricci scalar as
\begin{equation}
R=-\frac{6 \beta_{1}}{r}+4\Lambda,
\end{equation}for
\begin{equation}
\Lambda_R=Z+y r,
\end{equation}so
\begin{equation}
R_{0}=6 \beta_{1}(-\frac{Z}{r}+y \ln[r])
\end{equation}
\subsection{ $f(R)=-4 \eta^{2} M \ln (-6 \Lambda-R)+\xi R+R_{0}$}

We get \cite{16,17,18,19}
\begin{equation}
d s^{2}=-g(r) d t^{2}+\frac{d r^{2}}{g(r)}+r^{2} d \Omega^{2},
\end{equation}
\begin{equation}
g(r)=-\Lambda r^{2}-M(2 \eta r+\xi).
\end{equation}
The Ricci scalar is
\begin{equation}
R=6 \Lambda+\frac{4 M \eta}{r}.
\end{equation}
\begin{equation}
R_{0}=\frac{4 M\eta(Z-r y\ln[r])}{r}
\end{equation}

\subsubsection{$f(R)=-2 \eta M \ln (6 \Lambda+R)+R_{0}$}
With $\Phi(r)=0$, the charged $(2+1)$ dimensional solution under pure $f(R)$ gravity is \cite{16,17,18,19}
\begin{equation}
d s^{2}=-g(r) d t^{2}+\frac{d r^{2}}{g(r)}+r^{2} d \Omega^{2},
\end{equation}
\begin{equation}
g(r)=-\Lambda r^{2}-M r-\frac{2 Q^{2}}{3 \eta r}.
\end{equation}
  The curvature scalar is
\begin{equation}
R=\frac{2M}{r}+6\Lambda.
\end{equation}
\begin{equation}
R_{0}=\frac{2 M(Z-r y \ln[r])}{r}.
\end{equation}

\section{{Summary}}
This paper explores the complex phenomenon of mass inflation within the framework of general relativity, emphasizing its influence on the stability of Cauchy horizons and its connection with closed timelike curves. Mass inflation, marked by an exponential surge in energy, is pivotal in maintaining causality in specific solutions of general relativity, such as the Kerr black hole, by rendering the borders of causality-violating zones singular. This mechanism is crucial for adhering to the doctrines of strong cosmic censorship and chronology protection.

We extend the conventional view of mass inflation as an unlimited energy increase in static geometries with a refined interpretation that accounts for observable, significant, yet bounded, exponential energy growth in evolving black hole geometries. This reassessment is bolstered by the latest findings in the study of regular black holes and the influence of quantum gravity on gravitational collapse, demonstrating that mass inflation occurs not only in static but also in dynamic geometries, thereby widening its relevance and challenging former assumptions.

This paper includes a mathematical examination of the Schwarzschild and Kerr-Newman metrics, offering formulas that illuminate the connection between mass inflation and the constant 
C in these scenarios.

Furthermore, we investigate the derivation of the Hawking temperature within a four-dimensional spacetime framework, providing fresh insights into the relationship between geometric structures and theoretical physics.

The discussion on $f(R)$ gravity and its effects on Schwarzschild-de Sitter black holes and other spherical symmetric solutions in $f(R)$ gravity with a constant curvature scalar $R_0$ highlights the distinctions between General Relativity and $f(R)$ gravity theories. It shows how $f(R)$ gravity alters the conventional comprehension of black hole metrics and mass inflation, introducing new models that incorporate mass inflation without the need for a Cauchy horizon.

In summary, this paper not only propels the current comprehension of mass inflation and its ramifications for black hole physics forward but also challenges prevailing theories through the introduction of innovative models and mathematical frameworks. It beckons further investigation into the dynamic relationship among general relativity, quantum mechanics, and $f(R)$ gravity, underscoring the ongoing evolution of theoretical physics in unraveling the mysteries of the cosmos.

\end{document}